\documentclass[%
 reprint,
 amsmath,amssymb,
 aps,
 prl,
 reprint
]{revtex4-1}

\usepackage{graphicx}% Include figure files
\usepackage{dcolumn}% Align table columns on decimal point
\usepackage{bm}% bold math
\usepackage{amsfonts}
\usepackage{fancyhdr}
\usepackage{notes2bib}
\begin{document}
\pagestyle{empty}
%\preprint{APS/123-QED}

\title{Bloch-Floquet waves in optical ring resonators}% Force line breaks with \\

\author{Kathleen McGarvey-Lechable}
 \author{Pablo Bianucci}%
 \email{pablo.bianucci@concordia.ca}
\affiliation{%
 Department of Physics, Concordia University, Montreal, QC, H4B 2J1, Canada}%

\date{\today}% It is always \today, today,

\begin{abstract}
Modal coupling between frequency-degenerate resonances of an optical ring resonator is a commonly observed phenomenon that results in adverse mode splitting. Traditionally, this coupling is attributed to Rayleigh scattering of a propagating electromagnetic wave into its associated degenerate counter-propagating mode from small perturbations to the dielectric material of the resonator.  We have chosen to reframe the problem of intracavity Rayleigh scattering by considering the optical ring resonator as an infinitely-long, one-dimensional photonic crystal (PhC) that possesses a lattice constant equal to the perimeter of the ring. Through application of Bloch-Floquet theory, we show that modal coupling between degenerate resonances of a ring can effectively be described as the formation of photonic frequency bands in the dispersion relation of the resonator. We additionally demonstrate that the Bragg planes of the PhC lattice coincide with the phase matching conditions for constructive interference in the ring. Finally, we show that the magnitude of frequency splitting of a particular resonance is proportional to its associated coefficient in the Fourier expansion of the ring's periodic dielectric function.

\end{abstract}

\pacs{42.25.-p,42.82.Et,78.67.Pt}% PACS, the Physics and Astronomy
%                              % Classification Scheme.
%                               %display desired
\maketitle

%\tableofcontents

\textit{Introduction.---} Optics and condensed matter physics have long been complementary fields, with advances in one inspiring new ideas in the other. In recent years, they have become further entwined in a new way, with inspiration from the field of topological condensed matter physics laying the foundations for the burgeoning field of topological photonics\cite{raghu2008quantumhall}. For instance, observation of protected edge states in arrays of optical microresonators\cite{hafezi2013imaging} and Weyl nodes in microwave photonic crystal slabs\cite{chen2016weylpoints} have been reported.% Any more examples?  
The key feature in these demonstrations is the presence of spatial periodicity, which allows for the use of Bloch theory in finding appropriate solutions to Maxwell's equations and gives a formal underpinning to the analogy.

Microscopic ring resonators (RRs) are a staple in the field of photonics, as a building block in integrated photonic circuitry\cite{almeida2004,bogaerts2004} and the substrate for many different ``on-chip" optical demonstrations\cite{razzari2010cmos, foster2011silicon, ferdous2011spectral, silverstone2015qubit, grassani2015micrometer,wakabayashi2015time, kues2017, feng2014, hodaei2014, hausmann2012}. A characteristic feature of RRs is the splitting of modal resonances due to scattering of a propagating wave into a frequency-degenerate counter-propagating one. Due to its implications for optical device performance\cite{Fujii:17}, modal coupling in ring resonators has been the subject of intensive study, having previously been described via several robust approaches including steady state loop equations \cite{little1997}, temporal Coupled Mode Theory \cite{little1997microring, zhang2008, wang2009, ballesteros2011, li:10, li2016}, and the transfer matrix method \cite{kang2010, gao2015}. While these approaches are very useful, they are usually limited to analyzing the splitting of a single mode. 

In this work, in order to tackle modal coupling in a more general way, we have chosen to map the optical RR to an equivalent condensed-matter inspired system. Small perturbations to the dielectric profile of the resonator are encountered repeatedly by an optical wave as it makes multiple circuits around the resonator. The perturbed dielectric profile of the ring can thus be modeled as a one-dimensional photonic crystal which possesses a lattice constant equal to the perimeter of the resonator.  Accordingly, direct parallels can be drawn between the optical spectrum of a ring resonator and the well-understood energy band structure of a 1-D solid state crystal in the presence of a weak atomic potential.  In particular, Bloch-Floquet theory can be used to effectively describe  the modal coupling of frequency-degenerate resonances in a ring as the formation of photonic bands in the resonator's dispersion relation.  This analysis can additionally be extended to engineer the splitting of specific ring resonances, obtaining spectral features that are immune to additional frequency splitting.

\textit{Theoretical Analysis---} The propagation of light through an isotropic, lossless dielectric material in the linear regime can effectively be described by solving the electromagnetics generalized  eigenvalue problem corresponding to a harmonic electric field with angular frequency $\omega$, $\textbf{E}(\textbf{r},t) = \textbf{E}(\textbf{r})e^{-i\omega t}$:
\begin{equation}
  \nabla \times \Big[ \nabla \times \textbf{E}(\textbf{r})\Big] =  {\bigg(\frac{\omega}{c}\bigg)}^2 \epsilon(\textbf{r}) \textbf{E}(\textbf{r}),
  \label{eq:emmaster}
\end{equation}
where $\epsilon(\textbf{r})$ is a function delineating the geometry of the dielectric material. Here we have assumed negligible material dispersion and transverse electromagnetic fields. 

When applied to a cylindrically-symmetric ring resonator, several assumptions can be made which simplify Eq. (\ref{eq:emmaster}). Provided the radius, $R$, of the resonator is large enough (i.e. $R >> \lambda_0/\sqrt{\epsilon(\textbf{r})}$ where $\lambda_0$ is the wavelength of interest), the resonator can be represented as an infinitely-long, two-dimensional waveguide of effective dielectric constant $\epsilon(\textbf{r}) = \epsilon(x) $.  In such a waveguide, light will effectively propagate along the $x$-axis, allowing us to express the wavevector of the electromagnetic field as $\textbf{k}(\textbf{r}) = k_x(x)\ \hat{\textbf{x}}= k(x)\ \hat{\textbf{x}}$.  Finally, we will consider only fields linearly polarized along the $y$-axis, reducing the electric field to $\textbf{E}(\textbf{r}) = E(x)\ \hat{\textbf{y}}$ . 
With these assumptions in place, application of a simple vector identity ($\nabla \times \nabla \times \boldsymbol{A} = \nabla(\nabla \cdot \boldsymbol{A}) - \nabla^2 \boldsymbol{A})$ transforms Eq. ({\ref{eq:emmaster}}) into the well known one-dimensional Helmholtz equation:
\begin{equation}
  \epsilon^{-1}(x)\frac{\partial^2 E}{\partial x^2}  = -{\bigg(\frac{\omega}{c}\bigg)}^2 E(x).
  \label{eq:emmaster1}
\end{equation}
This equation is equivalent in form to the time-independent Schrodinger's equation. As a result, parallels between the electromagnetic modes in a ring resonator and the allowed electron wavefunctions in a potential energy landscape can easily be drawn.  Specifically, the formation of frequency (energy) bands in the dispersion relation of a ring resonator (1D crystalline structure) can be attributed to discrete translational symmetry constraints due to the periodicity of the dielectric (potential energy) function of the system.

\textit{Optical ring resonator as a 1D photonic crystal.---} We have until now considered the ring resonator as an infinitely-long waveguide of dielectric constant, $\epsilon_0$.  Such a system possesses continuous translational symmetry, ensuring that for any given eigenstate with frequency eigenvalue $\omega$ and corresponding wave number $k$ there exists a frequency degenerate eigenstate with wavenumber $-k$.  These eigenstates possess equal and opposite group velocities (i.e. $\pm \frac{d\omega}{dk}$) and correspond to the propagating and counter-propagating degenerate modes of the system.

\begin{figure}
  \begin{center}
  \includegraphics[width=\columnwidth]{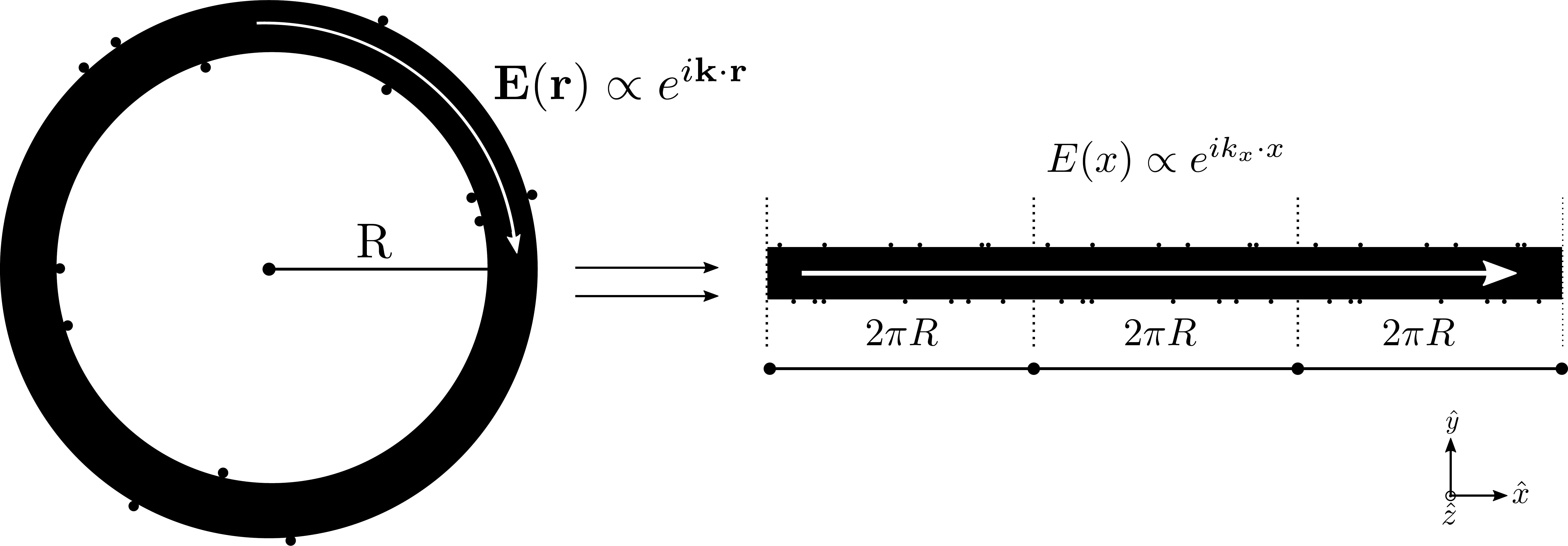}
  \caption{A standard ring resonator of radius $R$ with point surface scatterers represented as an infinitely-long dielectric waveguide.  The high dielectric constant material is seen in black, while the low dielectric constant material is seen in white.}
  \label{fig:RR1}
  \end{center}
\end{figure}

While in theory these degenerate modes can be excited in the ring resonator independently of one another, practical considerations such as fabrication-induced surface roughness and input-output coupling ports break the continuous translational symmetry of the system.  As a result, the dielectric function of the ring resonator forms a one-dimensional photonic crystal lattice, with a lattice constant equal to the perimeter of the ring, $P=2\pi R$ (see Figure \ref{fig:RR1}).
\begin{equation}
  \epsilon(x) = \epsilon(x + P) = \epsilon(x+ 2\pi R).
  \label{eq:periodice}
\end{equation}

\begin{figure}
  \centering
  \includegraphics[width=0.475\textwidth]{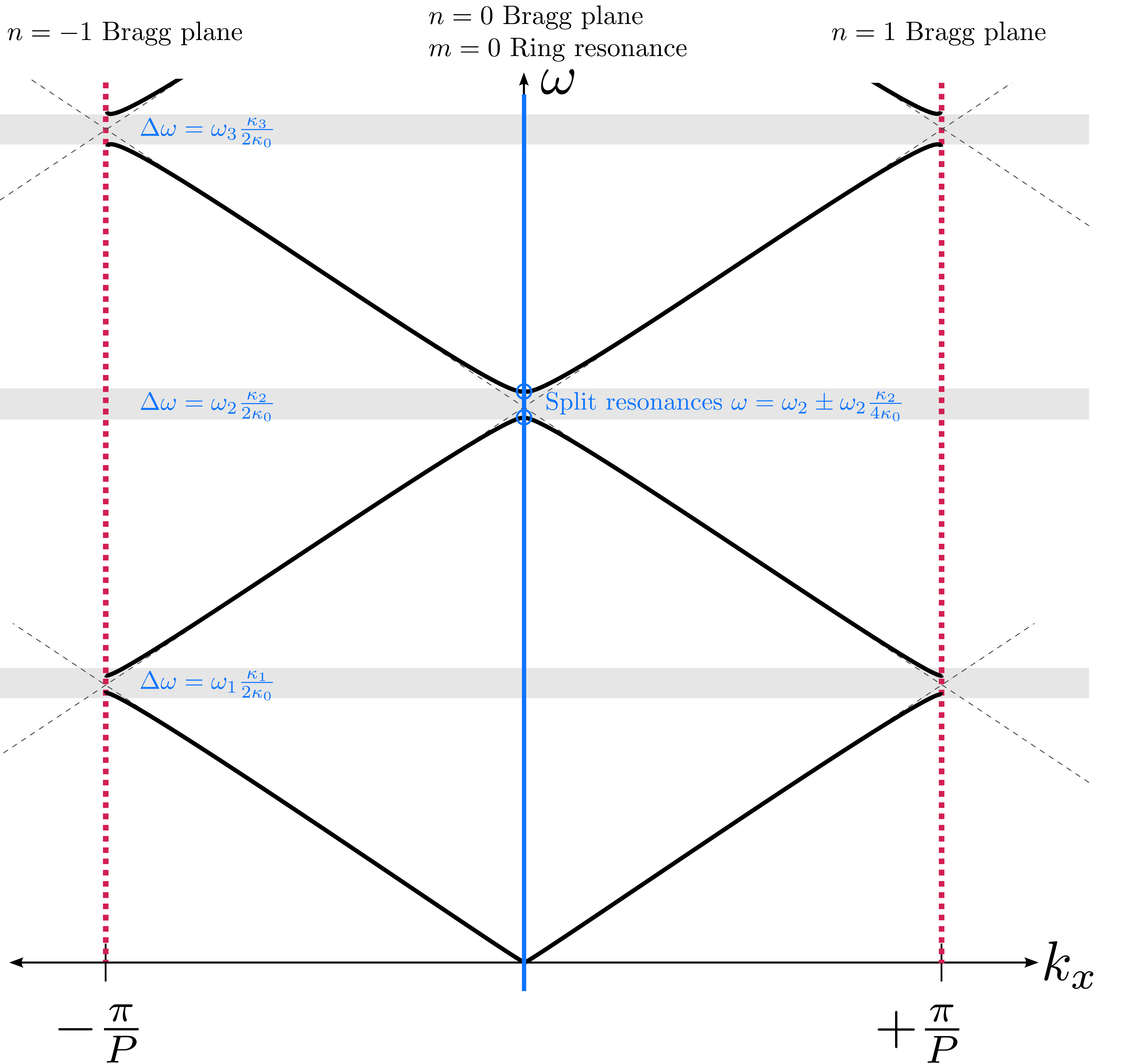}
  \caption{The photonic dispersion relation of a ring resonator in the reduced zone scheme (assuming a linear waveguide dispersion). The resonances of the ring occur wherever the phase matching boundary conditions (the horizontal lines) intersect the photonic dispersion relation of the infinitely-long waveguide. Phase matching conditions are separated by two reciprocal lattice vectors; consequently only the $m = 0$ boundary conditions represent physically distinct states in the reduced zone scheme. }
  \label{fig:RR2}
\end{figure}

Applying Bloch-Floquet theorem to Eq. (\ref{eq:emmaster1}), it can be shown \cite{sakoda_2005} that the electric field eigenfunctions of the system must likewise satisfy:
\begin{equation}
  E(x) = E(x+2\pi R).
  \label{eq:perwf}
\end{equation}
The periodicity of the resonator's dielectric profile thus results in the formation of Bragg planes at the points of symmetry in reciprocal space given by:
\begin{equation}
  k = \frac{n\pi}{P} =  \frac{n}{2R}, \quad \quad n \in \mathbb{Z}.
  \label{eq:phaseperiodic}
\end{equation}
Where these Bragg planes intersect the photonic dispersion relation of the 1D, infinitely-long waveguide, the frequency-degeneracy of the propagating and counter-propagating modes is lifted, resulting in the formation of photonic energy bands. The frequency-degenerate modes are effectively coupled, forming two new resonances of the system composed of linear combinations of the two original modes.   

The relation between the ring's photonic energy bands and its resonances is found by examining the phase matching conditions for an electromagnetic resonance.  A supported mode of the ring must possess a wavenumber which interferes with itself as it makes multiple round trips about the resonator:
\begin{equation}
  k = \frac{m}{R}, \quad \quad m \in \mathbb{Z}. 
  \label{eq:phasematching}
\end{equation}
Each one of these allowed wave vectors corresponds to a  resonance of the ring formed due to constructive interference of the electric field, resulting in amplification of the field strength. Combining Equations  (\ref{eq:phaseperiodic}) and (\ref{eq:phasematching}), it is evident that a  resonance of the ring coincides with the Bragg planes of the system whenever $n = \{2m:m \in \mathbb{Z}\}$ (see Figure \ref{fig:RR2}). Consequently, modal coupling resulting in frequency splitting of modes in ring resonators can be effectively described as the formation of photonic bands in the dispersion relation of the resonator due to discrete translational symmetry constraints. The next section will detail how Fourier analysis can be applied to calculate and engineer the magnitude of frequency splitting in ring resonators. 

\begin{figure}
  \centering
  \includegraphics[width=0.475\textwidth, clip]{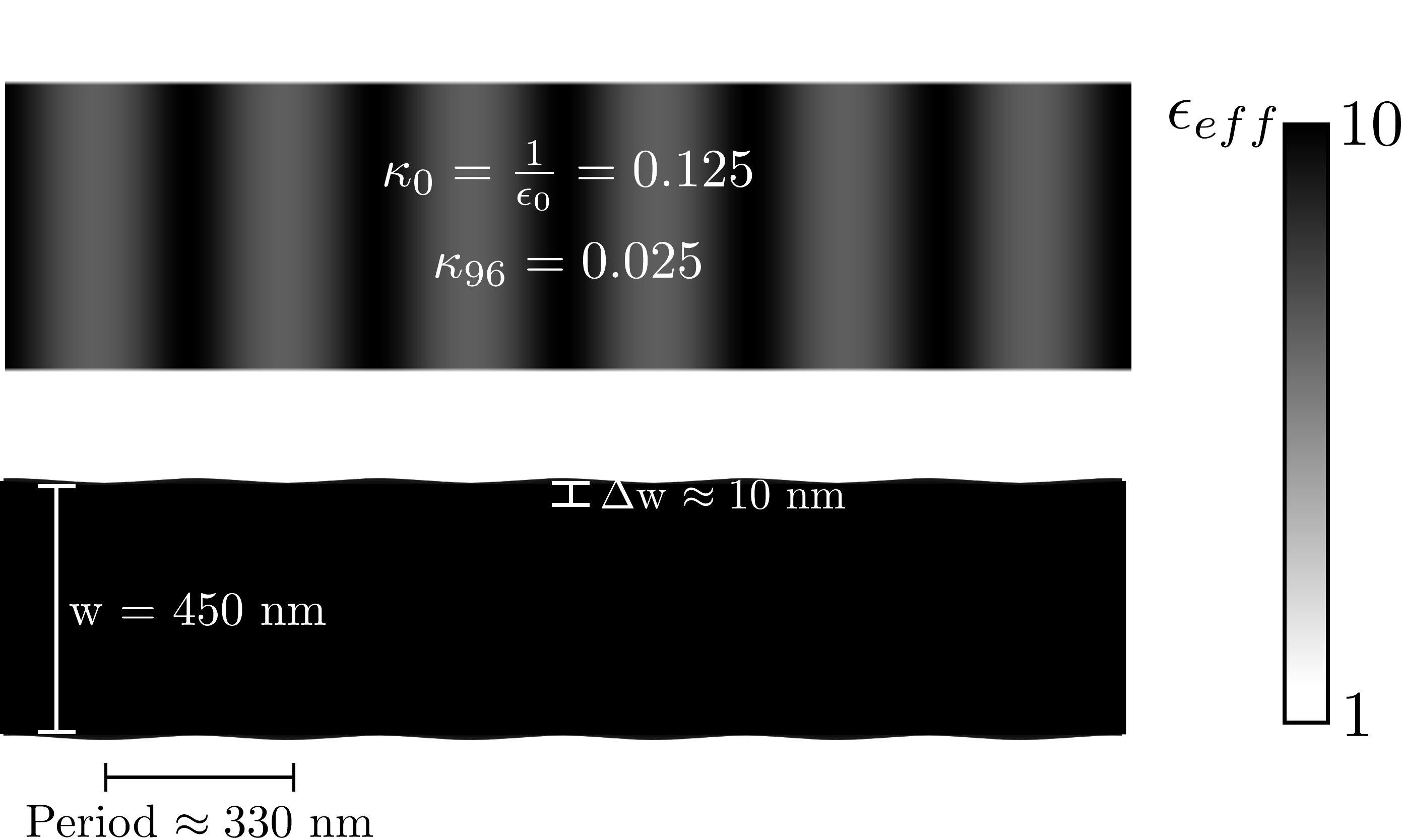}
  \caption{The dielectric function of a infinitely-long SOI waveguide of width $w =$ 450 nm and effective dielectric constant $\epsilon_0$ = 8.01.  The top image represents the Fourier expansion of the dielectric function of a waveguide with $\kappa_0 = 0.125$, $\kappa_{96} = 0.025$ and $\kappa_{l \neq 96} = 0$.  The bottom image shows the geometry of an equivalent waveguide.  For a $5-\mu$m-radius ring, the waveguide possesses a dielectric function modulation of wavelength $\lambda = \frac{2\pi R}{96} = 330$ nm and amplitude $\Delta w = 10$ nm.}
  \label{fig:eps}
\end{figure}
\textit{Quantitative Fourier analysis of mode splitting. ---} In choosing to consider the optical resonator as a one-dimensional photonic crystal, it is convenient to represent the inverse periodic dielectric material of the resonator as a Fourier series
\begin{equation}
  \epsilon^{-1}(x) = \sum_{l=-\infty}^{\infty} \kappa_le^{i2\pi lx/P},
  \label{eq:dielectricfourier}
\end{equation}
where $l \in \mathbb{Z}$ and $\kappa_l$ are the dielectric function's Fourier coefficients. Pursuant to Bloch-Floquet theorem, the electric fields satisfying Eq. (\ref{eq:emmaster}) can similarly be expressed as a Fourier series over the reciprocal lattice vectors of the Bravais lattice:
\begin{equation}
  E(x) = \sum_{l=-\infty}^\infty E_l e^{i(k+\frac{2\pi l}{P})x}.   
\end{equation}
Following the derivation detailed in Ref. \cite{sakoda_2005}, these two expressions can be inserted into the eigenvalue equation given by Eq. (\ref{eq:emmaster1}).  If we assume the $0^{th}$ and $\pm l$ terms dominate the Fourier expansion of the dielectric function, we can solve the resultant linear set of coupled equations to find the frequency eigenvalues of the periodic system.  Near the photonic Bragg planes (found at $k = \pm \frac{\pi l}{P}$), it can be shown that no resonance of the system can be found in the gap
\begin{equation}
  \Delta\omega = \omega_{l} \frac{\kappa_l}{2\kappa_0},
  \label{eq:deltaomega}
\end{equation}
where $\omega_l = \frac{l \sqrt{\kappa_0}}{2}\big(\frac{c}{P}\big)$ represents the unperturbed dimensionless frequency of the $l^{th}$-order mode of the ring resonator (see Supplementary Information). From this equation, it is evident that the magnitude of the frequency splitting of each resonance scales linearly with the unperturbed frequency of the resonance, $\omega_l$. This result is consistent with Rayleigh's classical scattering formula, which indicates elastic collisions between photons and dipole scatterers are enhanced at high frequencies. 

\begin{figure}
  \centering
  \includegraphics[width=3.25in, trim = 2cm 0 0 0, clip]{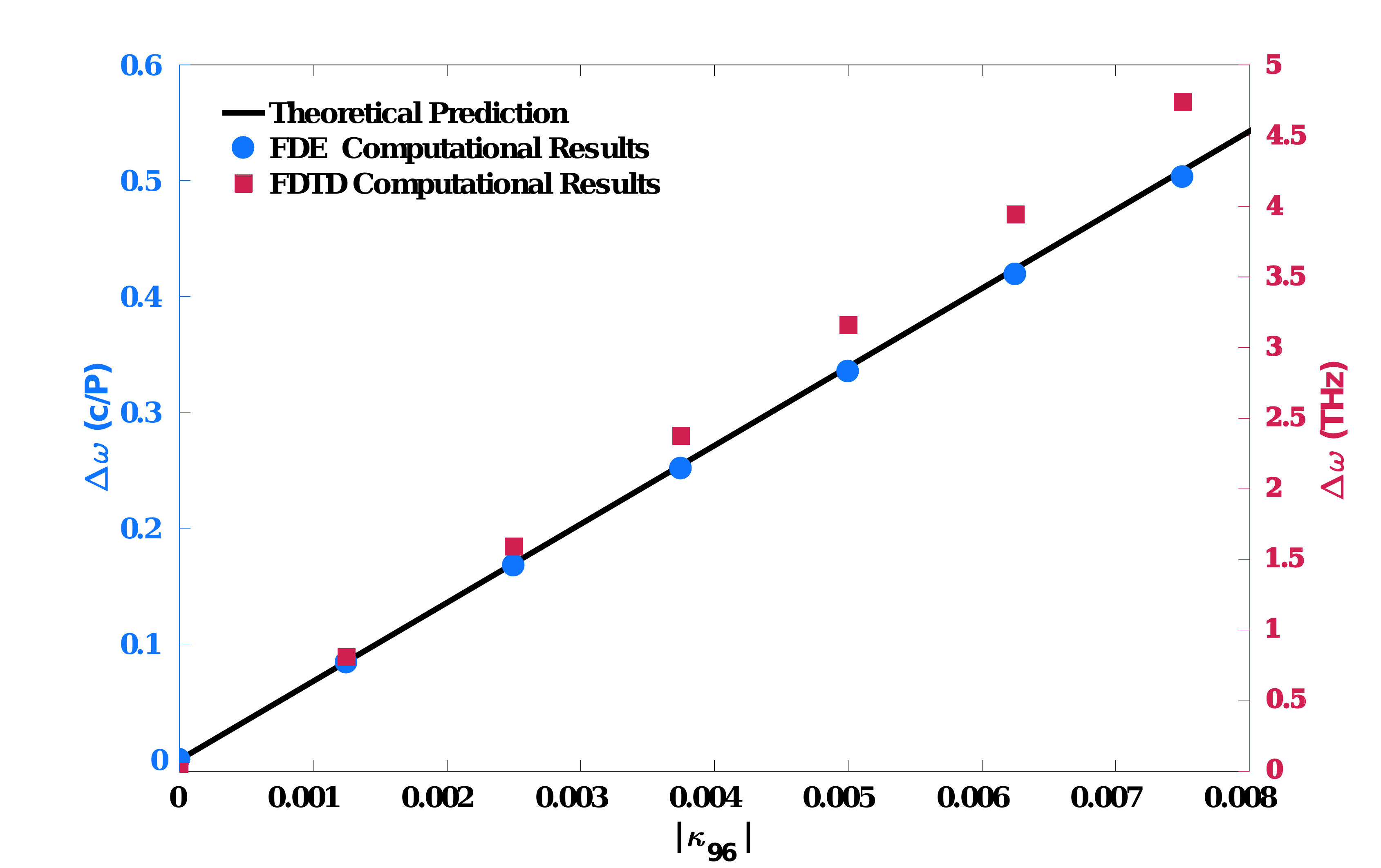}
  \caption{The numerical prediction of Eq. (\ref{eq:deltaomega}) as compared to FDE and FDTD computational results for various values of $\kappa_{96}$.  The dimensionless frequency of the FDE results has been scaled to the characteristic length unit for the system, the ring's perimeter $P$, while the FDTD results are given in THz.}
  \label{fig:k48}
\end{figure}

\textit{Frequency domain eigenmode solver (FDE) simulations.---}
A FDE solver \cite{johnson2001} is used to compute the effect of a small dielectric perturbation on the 1D photonic dispersion relation of an infinitely-long, 2D silicon-on-insulator (SOI) waveguide of width $w$ = 0.45 $\mu m$.  We have assumed a 220-nm-thick silicon layer on a silica substrate with air cladding and have chosen to consider only the fundamental, transverse-electric (TE) polarized mode, corresponding to a waveguide of effective dielectric constant $\epsilon_0 = 8.01$. Thanks to the periodicity imposed by the ring, we can expand the inverse dielectric function of the waveguide into a Fourier series:
\begin{equation}
  \epsilon^{-1}(x) = \kappa_0 + \sum_{2l:l \in \mathbb{Z}}\kappa_l  \cos \bigg(\frac{2\pi lx}{P}\bigg),
\end{equation}
where $\kappa_0 = \frac{1}{\epsilon_0}$. Because we have assumed a lossless material, we have kept only the real part of the inverse dielectric function defined in Eq. (\ref{eq:dielectricfourier}).  Additionally, the Fourier expansion is independent of frequency and thus does not account for waveguide or material dispersion. % This could be written better, but I'm not sure how right now.

As an example, we have chosen to perturb a resonance of frequency 196 THz (1530 nm) of a 5-$\mu m$-radius ring.  This resonance falls within the telecom C band and is thus highly suitable for applications in optical communications.  In order to induce modal coupling of this particular resonance, the Fourier coefficient corresponding to the resonance's mode number, $m$, must be selectively turned on. With dimensionless frequency units of ($\frac{2\pi c}{a}$) where $a$ = 0.45 $\mu m$ \bibnote{The FDE solver requires a length unit declaration with which to scale the photonic dispersion relation. While traditionally the photonic crystal's lattice constant is used (i.e. $P = 2\pi R$), better computational grid resolution can be achieved by using a smaller length unit (e.g. $a$ where $2\pi R = Na$ for $N \in \mathbb{Z}$). Here we have chosen to define the waveguide width, $w = 0.45$ $\mu m$,  as the arbitrary lattice constant $a$.  A 5-$\mu m$-radius ring can thus be simulated as a supercell photonic crystal of length $Na$ where $N=70$.}, the mode number of the resonance in question is $m = 96$, implying that the 196 THz  mode  can be split by modulating the Fourier coefficient $\kappa_{96}$.  The resulting one-dimensional dielectric function, and the waveguide corresponding to it can be seen in Figure \ref{fig:eps}. Figure \ref{fig:k48} compares the prediction of Equation (\ref{eq:deltaomega}) to the computational results from the frequency-domain eigensolver, showing excellent agreement. Additionally, negligible frequency splitting is observed in the remaining resonances of the ring, indicating that an $m^{th}$-order resonance remains unaffected by an $\{l \neq m\}$-order perturbation to the dielectric function. 

\begin{figure}
  \centering
  \includegraphics[width=0.475\textwidth]{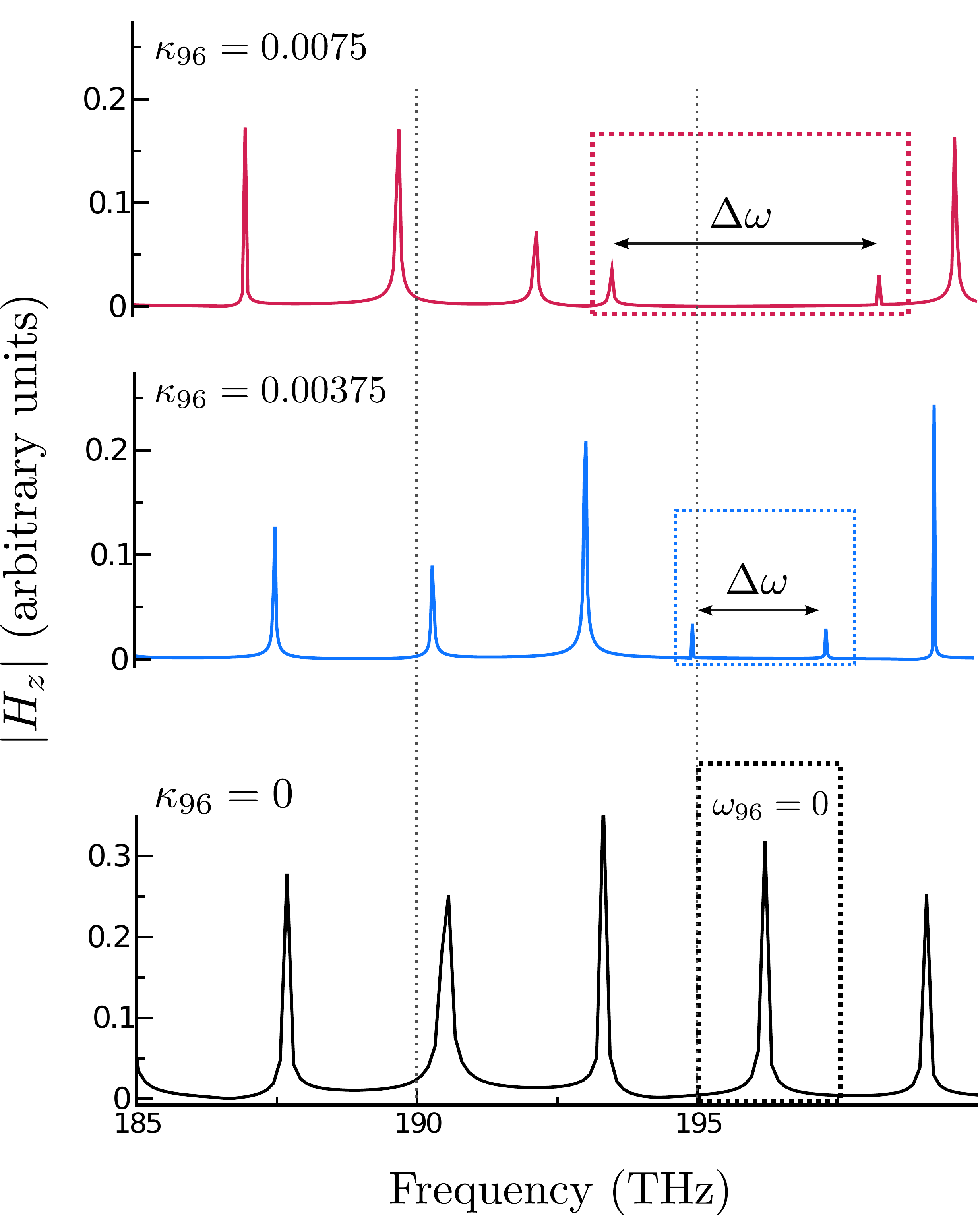}
\caption{The frequency spectrum of a $5-\mu$m 2D SOI ring resonator for $\kappa_{96} =$ 0, 0.00375, and 0.0075}
  \label{fig:spectrum}
\end{figure}

\textit{Finite-difference-time-domain (FDTD) simulations.---}
FDTD simulations \cite{oskooi2010} of a 2D SOI ring resonator of radius $5$ $\mu m$, width $w = 0.45$ $\mu m$ and varying magnitudes of $\kappa_{96}$ are also compared. We have again considered only the fundamental TE mode of the ring and have excited the electric fields using a broadband Gaussian dipole current source centered on 196 THz.   Figure \ref{fig:spectrum} displays these results, where the optical spectrum are obtained by taking the Fourier transform of the post-source electric fields.  

% We are not referring at all to this figure in the text, so we should take it out. That helps us with the length, too! If needed, we can discuss this in the supplementary material.
%\begin{figure}[t!]
%  \centering
%  \includegraphics[width=0.475\textwidth]{meepeps.eps}
%  \caption{The $H_z$ field configurations of the split resonances of a  5-$\mu$m-radius SOI 2D ring resonator with a periodic modulation of the dielectric function of magnitude $\kappa_{96}$ = 0.00625 and wavelength $\lambda \approx $ 330 nm. The two orthogonal resonances are offset from one another by a phase of $\delta = \frac{2\pi}{96*2}$.}
%  \label{fig:meepeps}
%\end{figure}

The unperturbed resonances  of order $l \neq 96$ show no significant mode splitting. The slight  shift of all resonances to the red with increasing perturbation can be attributed to a small increase in the effective index of the dielectric material, $\epsilon_0$. The black dotted box shows the resonance of interest, $\omega_{96} = 196$ THz.  When $\kappa_{96} = 0$, no modal coupling is observed in the system.  However, the rings with $\kappa_{96} \neq 0$ demonstrate significant frequency splitting of the resonance $\omega_{96}$, proportional in magnitude to $\kappa_{96}$ (see Figure \ref{fig:k48}). The slight variation of the FDTD simulation results from the prediction of Eq. (\ref{eq:deltaomega}) can be attributed to the additional perturbation to the system presented by the ring's radius of curvature.  Furthermore, because the coupled resonances of the system respect the discrete translational symmetry constraints of the perturbed ring, they are immune to further modal coupling or frequency-splitting; any additional perturbation to the system will merely contribute to increasing the magnitude of the splittings between the modes.

\textit{Conclusions.---}  In summary, we have reformulated the problem of frequency splitting in ring resonators by considering the ring as a one-dimensional PhC and applying a band structure analysis.  Since the system has a well-defined periodicity, Bragg planes appear in reciprocal space. We find that some of these Bragg planes coincide with the phase matching conditions for constructive interference in the ring. A perturbation approach, similar to the one used for nearly-free electrons in a one-dimensional atom chain, yields that the magnitude of modal coupling of the $l$-th resonance of a ring is proportional to its corresponding coefficient of the dielectric function's Fourier expansion. Subsequently, specific resonances of the ring can be designedly split by precisely engineering periodic modulations of the ring.  The resultant modes respect the discrete translational symmetry constraints of the system and thus are immune to further frequency splitting.

\newpage
\section{Supplementary Information}
\textit{Derivation of frequency mode splitting. --}  As seen in Eq. \ref{eq:periodice}, the material of the ring resonator can be considered periodic, allowing for the expression of the ring's dielectric function as a Fourier series:

\begin{equation}
\epsilon^{-1}(x) = \sum_{m = - \infty}^{\infty}\kappa_m e^{i2\pi mx/P} \quad \quad m \in \mathbb{Z}
\label{eq:sup1}
\end{equation}

The allowed electric fields in the resonator can similarly be expressed as a Fourier series:
 
\begin{equation}
E(x) = \sum_{m = -\infty}^\infty E_m e^{i(k+\frac{2\pi m}{P})x}.
\label{eq:sup2}
\end{equation}

We now assume that the Fourier expansion of the dielectric function of the ring is dominated by the $m = 0$ and $m = \pm l$ components:

\begin{equation}
\epsilon^{-1}(x) = \kappa_0 + \kappa_le^{i2\pi lx/P} + \kappa_{\text{-}l}e^{-i2\pi lx/P}.
\label{eq:sup3}
\end{equation}

Combining Eqs. \ref{eq:sup2} and \ref{eq:sup3} with the 1D electromagnetics wave equation (Eq. \ref{eq:emmaster1}), we find:

\begin{multline*}
  -\sum_{j=-1}^{1} \sum_{m = -\infty}^{\infty} \kappa_{l \times j} e^{i2\pi(l \times j)x/P}E_m e^{i(k+\frac{2\pi m}{P})x}\bigg(k + \frac{2\pi m}{P} \bigg)^2
  \\ = -\Big(\frac{\omega}{c}\Big)^2\sum_{m = -\infty}^{\infty}E_m e^{i(k+ \frac{2\pi m}{P})x} 
\end{multline*}
\begin{multline*}
  \Rightarrow -\sum_{m = -\infty}^\infty \kappa_{\text{-}l}E_{m+l}e^{-i2\pi l x/P} \; e^{i\frac{2\pi(m+l)}{P}x}\bigg[k+\frac{2\pi(m+l)}{P}\bigg]^2 \\
  - \kappa_l E_{m - l} e^{i2\pi lx/P} \; e^{i\frac{2\pi(m-l)}{P}x}\bigg[k+\frac{2\pi(m-l)}{P}\bigg]^2\\
  =\sum_{m = -\infty}^\infty\bigg[\kappa_0\Big(k + \frac{2\pi m}{P}\Big)^2 - \Big(\frac{\omega}{c}\Big)^2\bigg]E_m e^{i\frac{2\pi m}{P}x}
\end{multline*}

\begin{multline}
  \Rightarrow \sum_{m = -\infty}^\infty\kappa_{\text{-}l}\bigg[k + \frac{2\pi(m+l)}{P}\bigg]^2E_{m+l} \\+ \kappa_l\bigg[k + \frac{2\pi(m-l)}{P}\bigg]^2E_{m-l} \\
  = \sum_{m = -\infty}^\infty \bigg[\Big(\frac{\omega}{c}\Big)^2 - \kappa_0\Big(k + \frac{2\pi m}{P}\Big)\bigg]^2E_m
\label{eq:sup4}
\end{multline}

When $m$ = 0, Eq. \ref{eq:sup4} becomes:
\begin{multline*}
  \kappa_{\text{-}l}\Big(k + \frac{2\pi l}{P}\Big)^2E_l \; + \; \kappa_l\Big(k - \frac{2\pi l}{P}\Big)E_{\text{-}l} \\
  = \bigg[\Big(\frac{\omega}{c}\Big)^2 - \kappa_0k^2\bigg]E_0\\
\end{multline*}
\begin{equation}
\Rightarrow E_0 \simeq \frac{1}{\big(\frac{\omega}{c}\big)^2 - \kappa_0k^2}\bigg[\kappa_{\text{-}l}\Big(k + \frac{2\pi l}{P}\Big)^2E_l + \kappa_l\Big(k-\frac{2\pi l}{P}\Big)^2E_{\text{-}l}\bigg]
\label{eq:sup5}
\end{equation}

Similarly, for $m = -l$ we find:
\begin{multline*}
  \kappa_{\text{-}l}k^2E_0 \; + \; \kappa_l\Big(k - \frac{4\pi l}{P}\Big)^2E_{\text{-}2l} \\
  = \bigg[\Big(\frac{\omega}{c}\Big)^2 - \kappa_0\Big(k - \frac{2\pi l}{P}\Big)^2\bigg]E_{\text{-}l}
\end{multline*}
\begin{equation}
  \Rightarrow E_{\text{-}l} \simeq \frac{1}{\big(\frac{\omega}{c}\big)^2-\kappa_0\big(k - \frac{2\pi l}{P}\big)^2}\bigg[\kappa_l\Big(k - \frac{4\pi l}{P}\Big)^2E_{\text{-}2l}\; + \; \kappa_{\text{-}l}k^2E_0\bigg]
\label{eq:sup6}
\end{equation}
Provided the dispersion relation is approximately linear (i.e. $\omega \simeq c\sqrt{\kappa_0}k$), then these two terms will dominate the expansion whenever $k \simeq \pm \frac{2\pi l}{P}$.  As a result, Eq. \ref{eq:emmaster1} is reduced to a linear set of coupled equations:

\begin{align}
  \begin{split}
    \bigg[\Big(\frac{\omega}{c}\Big)^2 - \kappa_0k^2\bigg]E_0 - \kappa_l\Big(k - \frac{2\pi l}{P}\Big)^2E_{\text{-}l} & = 0 \\
    \kappa_{\text{-}l}k^2E_0 + \bigg[\Big(\frac{\omega}{c}\Big)^2- \kappa_0\Big(k - \frac{2\pi l}{P}\Big)\bigg]^2E_{\text{-}l} &= 0
  \end{split}
\end{align}
Solutions to these coupled equations can be found by taking the determinant of the resultant coefficient matrix:

\begin{equation}
\begin{vmatrix}
\big(\frac{\omega}{c}\big)^2 - \kappa_0k^2 & -\kappa_l\big(k-\frac{2\pi l}{P}\big)^2\\
-\kappa_{\text{-}l}k^2 & \big(\frac{\omega}{c}\big)^2 - \kappa_0\big(k - \frac{2\pi l}{P}\big)^2
\end{vmatrix} = 0
\end{equation}
We now make a change of variables, setting $h = k - \frac{\pi l}{P}$:

\begin{equation*}
\begin{vmatrix}
\big(\frac{\omega}{c}\big)^2 - \kappa_0 \big(h + \frac{\pi l}{P}\big)^2 & -\kappa_l\big(h-\frac{\pi l}{P}\big)^2\\
-\kappa_{\text{-}l}\big(h + \frac{\pi l}{P}\big)^2 & \big(\frac{\omega}{c}\big)^2 - \kappa_0\big(h - \frac{\pi l}{P}\big)^2
\end{vmatrix} = 0
\end{equation*}
\begin{multline}
\Rightarrow \bigg[\Big(\frac{\omega}{c}\Big)^2 - \kappa_0\Big(h + \frac{\pi l}{P}\Big)^2\bigg]\bigg[\Big(\frac{\omega}{c}\Big)^2 - \kappa_0\Big(h-\frac{\pi l}{P}\Big)^2\bigg] \\
- \kappa_l^2 \Big(h + \frac{\pi l}{P}\Big)^2\Big(h - \frac{\pi l}{P}\Big)^2 = 0 
\label{eq:sup7}
\end{multline}
We have assumed the dispersion relation is approximately linear, implying that we will consider only the terms of Eq. \ref{eq:sup7} which are $\propto h^0$ and $h^4$:

\begin{equation}
\Big(\frac{\omega}{c}\Big)^4 \simeq h^4\big(\kappa_0^2 - \kappa_l^2\big) + \frac{\pi^4l^4}{16P^4}\big(\kappa_0^2 - \kappa_l^2\big)
\label{eq:sup8}
\end{equation}
We wish to determine the frequency eigenvalues near the photonic band edge (i.e. $h \simeq 0$).  At these points in reciprocal space, the dispersion relation is entirely determined by the intercept of Eq. \ref{eq:sup8}:

\begin{equation*}
\Big(\frac{\omega}{c}\Big)^4 \simeq \frac{\pi^4l^4\kappa_0^2}{16P^4}\Big(1 - \frac{\kappa_l^2}{\kappa_0^2}\Big) 
\end{equation*}
\begin{equation}
\Rightarrow \omega ^2  \simeq \pm\frac{{\omega_l}^2}{4} \sqrt{1 - \frac{\kappa_l^2}{\kappa_0^2}}
\label{eq:sup9}
\end{equation}
where $\omega_l \equiv \frac{cl\sqrt{\kappa_0}}{2P} \hspace{.2cm} (s^{-1}) = \frac{l\sqrt{\kappa_0}}{2}\big(\frac{c}{P}\big)$. If we assume $\kappa_l \ll \kappa_0$, a Taylor expansion of Eq. \ref{eq:sup9} yields:

\begin{equation}
  \omega^2=\begin{cases}
    +\frac{\omega_l^2}{4}\Big(1\pm\frac{\kappa_l^2}{2\kappa_0^2}\Big), & \text{positive roots}\\
     -\frac{\omega_l^2}{4}\Big(1\pm\frac{\kappa_l^2}{2\kappa_0^2}\Big), & \text{negative roots}
  \end{cases}
  \label{eq:omega2}
\end{equation}

As we are only interested in the real eigenvalues of the system, we will consider
only the positive roots of Eq. (\ref{eq:omega2}).  Solving for $\Delta\omega$ we find:

\begin{align}
\begin{split}
\Delta\omega^2 &= \frac{\omega_l^2}{4} + \frac{1}{2}\Big(\frac{\omega_l\kappa_l}{2\kappa_0}\Big)^2 - \frac{\omega_l^2}{4} - ^-\frac{1}{2}\Big(\frac{\omega_l\kappa_l}{2\kappa_0}\Big)^2 \\
&= \Big(\frac{\omega_l\kappa_l}{2\kappa_0}\Big)^2
\end{split}
\end{align}
From this equation, it is evident that no resonance of the ring resonator can be found in the gap:
\begin{equation}
\Delta \omega = \omega_l\frac{\kappa_l}{2\kappa_0} \hspace{0.3cm}
\end{equation}

\end{document}